\documentstyle[aps,prb,multicol,epsf]{revtex}

\draft
\begin{document}
\title{Nonadibaticity and single electron transport driven by surface acoustic waves}
\author{Karsten Flensberg,$^{1}$ Qian Niu,$^{2}$ and Michael Pustilnik$^{1,3}$}
\address{$^{1}$\O rsted Laboratory, Niels Bohr Institute for APG, \\
Universitetsparken 5, DK-2100 Copenhagen,\ Denmark\\
$^{2}$Department of Physics, The University of Texas at Austin, Austin, TX 78712, USA \\
$^{3}$Danish Institute of Fundamental Metrology, \\
Anker Engelunds Vej 1, Building 307, DK-2800 Lyngby, Denmark}
\date{\today}
\maketitle

\begin{abstract}
Single-electron transport driven by surface acoustic waves (SAW) through a
narrow constriction, formed in two-dimensional electron gas, is studied
theoretically. Due to long-range Coulomb interaction, the tunneling coupling
between the electron gas and the moving minimum of the SAW-induced potential
rapidly decays with time. As a result, nonadiabaticiy sets a limit for the
accuracy of the quantization of acoustoelectric current.
\end{abstract}


\begin{multicols}{2}
Recently, a new type of single electron devices was introduced. In the
experiments \cite{experiments}, surface acoustic waves (SAW) induce, via
piezo-electric coupling, charge transport through a point contact in GaAs
heterostructure. When the point contact is biased beyond the pinch-off, the
acoustoelectric current develops plateaus, where 
\begin{equation}
I=N_{0}ef.  \label{quantization}
\end{equation}
Here $f$ is SAW frequency, and $N_{0}$ is an integer. The plateaus were
shown to be stable over a range of temperature, gate voltages, SAW power,
and source-drain voltages. Remarkably high accuracy of the quantization (\ref
{quantization}), and high frequency of operation ($f\sim 3\;{\rm GHz}$)
immediately suggest a possibility of metrological applications of the effect 
\cite{review}.

However, deep understanding of these results is still lacking.
Qualitatively, the effect is explained by a simple picture of moving quantum
dots \cite{experiments}. Electrons, captured in the local potential minima
('dots'), created by SAW, are dragged through the potential barrier. The
strong Coulomb repulsion prevents excess occupation of the dot. Increase of
the SAW power deepens the dots, more states become available for the
electrons to occupy, and new plateaus appear. By changing the gate voltage,
the slope of the potential barrier can be lowered, which has a similar
effect.

Interestingly, the quantization was not observed in the open channel regime 
\cite{openchannel}, although it should be expected on the quite general
theoretical ground \cite{thouless}. For the mechanism of the quantization,
discussed in \cite{thouless}, it is essential that the DC conductance for
each instantaneous configuration of the SAW-induced potential is zero. In
the open channel regime it would require the channel length to be much
longer than SAW wavelength $\lambda $, which is difficult to realize. In the
experiments \cite{experiments} this problem is avoided, since in the
pinch-off regime the DC conductance is zero. However, as explained below,
the rapid change of SAW potential near the entrance to the channel creates a
new trouble, leading to nonadiabatic corrections to (\ref{quantization}).

\begin{figure}[tbp]
\centerline{\epsfxsize=4.5cm \epsfbox{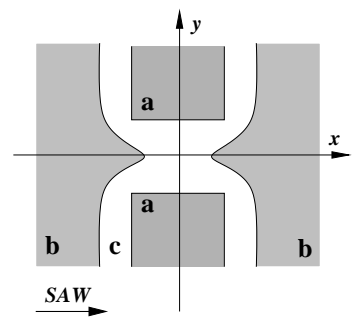}\vspace{0.3cm}}
\caption{ 
Geometry of the system: (a) split gates; (b) regions, occupied by 
two-dimensional electron gas; (c) depleted region. The arrow indicates 
direction of SAW propagation.
}
\label{Fig1}
\end{figure}

Long-range Coulomb interaction plays crucial role in this
phenomenon. The two-dimensional electron gas (2DEG) is depleted in the
vicinity of the gates (see Fig.1). On the other hand, in the depleted region
screening is lacking. The important parameters of the problem can be
understood from the solution of the electrostatic model. In this approach,
one assumes that, since the Fermi velocity $v_{F}$ is large compared to the
sound velocity $v_{s}$, the 2DEG is able to follow the changing in time
SAW-induced potential. Therefore, it is sufficient to consider an
instantaneous electrostatic problem, treating time as a parameter. Since the
screening length in 2DEG $(\sim 10\,{\rm nm})$ is much smaller, than $%
\lambda \sim 10^{3}\,{\rm nm}$, one can assume that the SAW-induced
potential is completely screened in the 2DEG-occupied region. Therefore, one
has to solve the Poisson equation subjected to complicated boundary
conditions. In particular, the potential at the gates ($\varphi =V_{g}$), as
well as the potential of 2DEG region ($\varphi =0$), and density $\rho =\rho
_{0}$ of the positive background charge in the depleted region are fixed.
Furthermore, for simplicity, one can take an effect of SAW into account
through a weak periodic modulation of $\rho $: $\rho \rightarrow \rho
_{0}+\delta \rho \left( x,t\right) $. The self-consistent solution would
yield the location of the edge of 2DEG, potential in the depleted region $%
\varphi \left( {\bf r}\right) $, and number density $n\left( {\bf r}\right) $
of 2DEG. Then, using the Thomas-Fermi-type relation $U\left( {\bf r}\right)
+\left( \pi \hbar ^{2}/m^{\ast }\right) n\left( {\bf r}\right) =\epsilon _{F}
$, one would be able to find an effective confining potential $U\left( {\bf r%
}\right) $. However,
the details of the full solution are not required, since the most important
properties can be understood from the following simple arguments 
\cite{conform}. 

Firstly, it is clear that the potential $\varphi \left( x,y\right) $ has a
minimum in the gap between the gates (see Fig.1). Secondly, the very
presence of plateaus shows that the charge states of the dot are separated by
the finite energy gaps. Since both $N_{0}={\rm odd}$ and 
$N_{0}={\rm even}$ plateaus are observed the energy gap is associated with
Coulomb repulsion rather than single-particle level spacing. 
This means that for any given $x$, $\varphi \left( x,y\right) $ has a sharp
minimum near $y=0$. The shape of this minimum is slowly changing with $x$,
and this change is controlled by the geometry of the device. Note that
the smoothness of the change of the confining potential in $y$-direction is supported by
the fact that the same systems exhibit very nice pattern of conductance
quantization in the open channel regime \cite{experiments}. Finally, the
weak perturbation of the background charge density $\left| \delta \rho
\right| \ll \rho _{0}$ should not affect significantly the position $x_{0}$
of the 2DEG edge.

\begin{figure}[tbp]
\centerline{\epsfxsize=8cm \epsfbox{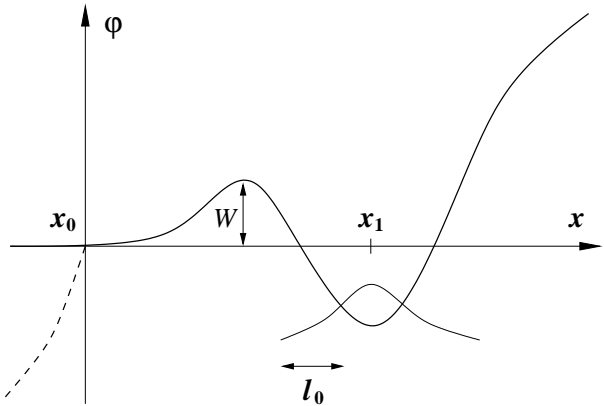}\vspace{0.3cm}}
\caption{ 
The potential $\varphi$ along the line $y=0$.  The dashed line shows the effective potential, that confines
2DEG to the region $x\leq x_0$. The SAW-induced 
potential minimum ($x_1$) moves to the right with the sound velocity $v_s$.
}
\label{Fig2}
\end{figure}

As function of time, $x_{0}\left( t\right) $ oscillates with SAW frequency $%
f $. However, the amplitude of the oscillations is small, compared to $%
\lambda $, therefore the velocity of the edge is negligible compared to $%
v_{s}$. On the other hand, the SAW-induced potential minimum (the dot) moves
away from the edge with precisely the sound velocity $v_{s}$ (see Fig.2).
Therefore, the width of the potential barrier, separating 2DEG and the dot,
increases linearly with time. This in turn means that the tunneling coupling
between 2DEG and the level, localized in the dot, rapidly decreases,
approximately exponentially. The characteristic time can be estimated as $%
\tau \sim l_{0}/v_{s}$, where $l_{0} $ is the distance over which the
localized wave extends under the barrier. Since, evidently, $l_{0}\ll
\lambda $, the relation 
\begin{equation}
f\tau \ll 1  \label{separation}
\end{equation}
holds. Other parameters, such as the height of the barrier $W$, the energy
of the localized level $\epsilon _{0}$, etc., change during the time, which
is of the order of SAW period $1/f$. Thus, the time dependence of all these
parameters during the time $\tau $ can be neglected.

Due to the rapid decrease of the tunneling coupling, the thermal equilibrium
in the system can not be maintained, causing fluctuations of the occupation
number of the dot. This results in nonadiabatic corrections to the quantized
values of the acoustoelectric current. Whether these corrections have a
significant impact on the accuracy of the quantization, depends on the value
of the characteristic energy scale $\hbar /\tau $, as compared to other
energy scales in the problem. To obtain an order of magnitude estimate, we
expand the potential near the minimum $x_{1}$ (see Fig.2), $V\left(
x\right) \sim Aq^{2}\left( x-x_{1}\right) ^{2}/2,\;q=2\pi /\lambda $. The
amplitude $A$ is related to the single particle level spacing in the dot $%
\Delta $ via $Aq^{2}=m^{\ast }\left( \Delta /\hbar \right) ^{2}$ ($m^{\ast }$
is effective electron mass), to the 'size' $r$ of the localized wave
function via $Aq^{2}r^{2}\sim \Delta $, and to the charging energy $E_{c}$
via $E_{c}\sim e^{2}/\epsilon r$. $l_{0}$ is estimated from WKB relation $%
\left( \hbar /l_{0}\right) ^{2}\sim 2m^{\ast }W$. Assuming that $W\sim A$,
it gives us four equations for five unknown quantities. Additional relation
follows from the experimental results. It was demonstrated \cite{experiments}%
, that the quantization disappears above the activation temperature $T^{\ast
}\sim 10\,{\rm K}$, which we identify with the charging energy. Using
typical parameters for the experiments \cite{experiments}, we find $\tau
\sim 10\,{\rm ps}$ \cite{nakamura}. All the parameters manage to pass minimal consistency
requirements $r\ll \lambda ,\;\Delta \ll A,E_{c},\;f\tau \ll 1$. Since the
corresponding energy scale $\hbar /\tau \sim 0.1\,{\rm meV}$, the
nonadiabatic effects may have significant influence on (\ref{quantization})
at low temperature.  

This can be understood from the following model Hamiltonian: 
\begin{equation}
H=H_{eg}+H_{dot}+H_{T}.  \label{model}
\end{equation}
Here 
\begin{equation}
H_{eg}=\sum_{k\sigma }\epsilon _{k}c_{k\sigma }^{\dagger }c_{k\sigma }
\label{2DEG}
\end{equation}
describes electron gas in the lead, 
\begin{equation}
H_{dot}=\sum_{n\sigma }E_{n}d_{n\sigma }^{\dagger }d_{n\sigma }+E_{c}\left(
N-{\cal N}_{g}\right) ^{2}  \label{dot}
\end{equation}
is the Hamiltonian of the dot ($N=\sum_{n\sigma }d_{n\sigma }^{\dagger
}d_{n\sigma }$ is the total number of electrons in the dot), and 
\begin{equation}
H_{T}=V(t)\sum_{kn\sigma }c_{k\sigma }^{\dagger }d_{n\sigma }+{\rm {H.c.}},
\label{tunneling}
\end{equation}
describes the tunneling coupling with time-dependent tunneling amplitude. We
have included only one lead in the model, since tunneling coupling to the
second lead is negligible. The electron gas in the lead is assumed to be in
thermodynamic equilibrium at all times, by virtue of the inequality $%
v_{F}\gg v_{s}$. As discussed above, we neglected time dependence of various
parameters in (\ref{dot}), due to the separation of the time scales $f\tau
\ll 1$. The last term in (\ref{dot}) describes the intra-dot Coulomb
interaction. The parameter ${\cal N}_{g}$ describes the effect of the gate
voltage. Since the width of the plateaus is approximately independent on the
plateau's number, it is a good approximation to assume that ${\cal N}$ is a
linear function of $V_{g}$. The most important ingredient of (\ref{tunneling}%
) is the time-dependent tunneling amplitude. We take it in the form 
\begin{equation}
V\left( t\right) =V_{0}e^{-t/\tau }  \label{V}
\end{equation}
(other possible choices will be discussed below). The time-independent
version of (\ref{model}-\ref{tunneling}) is commonly used in the theory of
the Coulomb blockade. Similar models have been also employed to study
transfer of charge during the atom-surface scattering \cite{AS}, and
nonadiabatic effects in charge pumping \cite{pump}.

Given that the system, described by (\ref{model}-\ref{tunneling}), is in
thermodynamic equilibrium at $t=-\infty $, our task is to calculate the
occupation of the dot at $t\rightarrow \infty $, 
$
N_{0}=\left\langle N\right\rangle _{t=\infty }  \label{No}
$
The acoustoelectric current is related to $N_{0}$ through (\ref{quantization}).

Away from the Coulomb blockade degeneracy points (half-integer ${\cal N}_{g}$),
when the inequality 
\begin{equation}
2E_{c}\left| {\cal N}_{g}-n_{0}-1/2\right| \gg \max \left\{ T,1/\tau \right\}
\label{ineq}
\end{equation}
is satisfied, the time-dependence is too slow to cause the transitions
between different charge states of the dot. In (\ref{ineq}), $n_{0}$ is
integer part of ${\cal N}_{g}$; units where $\hbar =k_{B}=1$ are used
throughout the rest of the paper. In this respect the evolution of the
system is almost adiabatic, and the occupation of the dot $N_{0}$ is
expected to coincide with equilibrium occupation, corresponding to the
Hamiltonian (\ref{dot}), with $T$ replaced by the effective temperature $%
T_{eff}\sim \max \left\{ T,1/\tau \right\} $. However, in the vicinity of
the transition region between the plateaus, when (\ref{ineq}) breaks down,
the time-dependence mixes states with $N=n_{0}$ and $N=n_{0}+1$. This means
that the width of the transition region is given by $T_{eff}$, and in the
zero-temperature limit saturates to $1/\tau $. In this regime the nearly
adiabatic picture fails. The width of the charge states due to tunneling $%
\Gamma (t)$ decreases with time. When $\Gamma (t)\lesssim 1/\tau $, the
system can no longer follow the changing tunneling coupling. Effectively, it
can be described within the sudden approximation, where $\Gamma (t)$ is
replaced by the step-function, 
$\Gamma (t)\rightarrow \Gamma _{s}\theta (-t),\;\Gamma _{s}\sim 1/\tau $ .
The occupation of the dot at $t\rightarrow \infty $ is therefore determined
by (\ref{model}-\ref{tunneling}) with {\it time-independent} tunneling
amplitude, corresponding to the width $\Gamma _{s}$.

Due to the interaction term in (\ref{dot}), the model (\ref{model}-\ref
{tunneling}) is still difficult to solve analytically. To simplify the
discussion, we limit our attention to the interval of ${\cal N}_{g}$, which
includes only one transition region between the plateaus: $n_{0}<{\cal N}%
_{g}<n_{0}+1$. Furthermore, we neglect the spin degeneracy, and consider the
limit of the large single-particle level spacing in the dot, 
$\Delta \gg 1/\tau$.
With these restrictions, at low temperature $T\ll \Delta $, only the lowest
energy configurations, corresponding to $N=n_{0}$ and $N=n_{0}+1$, are
important. Since these states are non-degenerate, one can introduce the
fermion operator $d=|n_{0}\rangle \langle n_{0}+1|$ to describe transitions between 
these states, 
and replace (\ref{dot}) by 
\begin{equation}
H_{dot}=E_{0}d^{\dagger }d,\;E_{0}=2E_{c}\left( 1/2+n_{0}-{\cal N}%
_{g}\right) .  \label{dot2states}
\end{equation}
The advantage of the model (\ref{model}),(\ref{tunneling}),(\ref{dot2states}%
) is that it is exactly solvable for arbitrary $V(t)$. Indeed, the
occupation of the dot $n\left( t\right) =\left\langle d^{\dagger
}(t)d(t)\right\rangle =N\left( t\right) -n_{0}\ $satisfies equation of
motion \cite{JWM} 
\begin{equation}
\frac{d}{dt}n(t)=-\Gamma (t)n(t)+\int d\varepsilon n_{F}(\varepsilon
)A\left( \varepsilon ,t\right) ,  \label{eom}
\end{equation}
\begin{equation}
A\left( \varepsilon ,t\right) =-\frac{1}{\pi }{\rm Im}\int dt^{\prime }\sqrt{%
\Gamma (t)\Gamma (t^{\prime })}e^{i\varepsilon (t-t^{\prime })}G^{R}\left(
t,t^{\prime }\right) .  \label{Jin}
\end{equation}
Here $\Gamma (t)=2\pi \nu V^{2}\left( t\right) $ is the width of the charge
state, $\nu $ is density of states of conduction electrons at the Fermi
level, $n_{F}(\varepsilon )$ is Fermi function and $G^{R}\left( t,t^{\prime
}\right) =-i\theta \left( t-t^{\prime }\right) \left\langle \left\{
d(t),d^{\dagger }(t^{\prime })\right\} \right\rangle $ is the exact retarded
Green function of the dot: 
\[
G^{R}\left( t,t^{\prime }\right) =-i\theta \left( t-t^{\prime }\right)
e^{-i\int_{t^{\prime }}^{t}dt_{1}\left[ E_{0}-i\Gamma \left( t_{1}\right)/2\right] }.
\]
Solution for $N_{0}$ follows by simple integration of (\ref{eom}-\ref{Jin}).
With $V\left( t\right) $ given by Eq. (\ref{V}), the result is 
\begin{equation}
N_{0}-n_{0}=\frac{\tau _{0}}{2\pi }\int_{-\infty }^{\infty }d\varepsilon 
\frac{n_{F}(\varepsilon )}{\cosh \left[ \left( \varepsilon -E_{0}\right)
\tau _{0}\right] },  \label{finiteT}
\end{equation}
where $\tau _{0}=\pi \tau /2$.
At zero temperature, (\ref{finiteT}) reduces to 
\begin{equation}
N_{0}-n_{0}=\frac{2}{\pi }\tan ^{-1}\left[ e^{-E_{0}\tau _{0}}\right] .
\label{zeroT}
\end{equation}
At finite temperature, (\ref{finiteT}) is described very well by the Fermi function 
\begin{equation}
N_{0}-n_{0}\approx \left( e^{E_{0}/T_{eff}}+1\right) ^{-1},  \label{Fermi}
\end{equation}
with an effective temperature $T_{eff}=\sqrt{\left( c/\tau _{0}\right)
^{2}+T^{2}}$. We found that $c=0.88$ gives the best numerical fit 
\cite{universality}.

For  $\left({\cal N}_{g}\rightarrow
n_{0}\right )$, Eqs. (\ref{quantization}) and (\ref{Fermi}) give the
following expression for the slope of the plateau: 
\begin{equation}
S=\frac{1}{I_{0}}\left( \frac{dI}{d{\cal N}_{g}}\right) _{{\cal N}%
_{g}\rightarrow n_{0}}\approx \left( 2E_{c}/T_{eff}\right)
e^{-E_{c}/T_{eff}}.  \label{slope}
\end{equation}
Here $I_{0}=n_{0}ef$ corresponds to perfect quantization. Strictly
speaking, to obtain the correct value of the slope precisely in the middle
of the plateau, the two-state approximation is not sufficient: state with $%
N=n_{0}-1$ makes exactly the same contribution, as that with $N=n_{0}+1$.
This complication, however, should not affect significantly the validity of (%
\ref{slope}): the exact result differs from (\ref{slope}) by the factor of
the order of $1$ only \cite{slope}. Due to the exponential factor in (\ref
{slope}), $S$ depends very strongly on the ratio $E_{c}/T_{eff}$. For
example, for $E_{c}/T_{eff}=10$, $S\sim 10^{-3}$, while for $E_{c}/T_{eff}=20
$, $S\sim 10^{-7}$.

According to the discussion above, in the transition region, the result can
be obtained with the sudden approximation. Thus, we have 
\[
N_{0}-n_{0} \approx \int d\varepsilon n_{F}(\varepsilon )\frac{\Gamma
_{s}/2\pi }{\left( \Gamma _{s}/2\right) ^{2}+(\varepsilon -E_{0})^{2}}, 
\]
or, for $T=0$, $N_{0}-n_{0} \approx 1/2-\left(1/\pi\right) \tan ^{-1}\left(2E_{0}/\Gamma _{s}\right)$.
This expression indeed coincides with (\ref{zeroT}) in the limit $%
E_{0}\rightarrow 0$, if $\Gamma _{s}/2=\tau _{0}^{-1}$. Note that the width
of the transition region is determined by essentially the same $T_{eff}$,
that enters (\ref{slope}).

The model (\ref{model}-\ref{tunneling}) introduced above allows study of the
nonadiabatic effects at the short time-scale. As the system evolves with
time, the SAW-induced potential minimum moves uphill (see Fig.2), and may
eventually cross the Fermi level. Due to the residual tunneling coupling in
this regime, the leakage from the dot will introduce additional corrections
to (\ref{quantization}). These corrections, however, do not affect strongly
the slope of the plateaus (\ref{slope}), and can be
taken into account by multiplying (\ref{quantization}) by the leakage factor 
$P_{l}\leq 1$: $I=P_{l}N_{0}ef$, where $P_{l}$ is expected to depend on
system parameters, such as gate voltage and SAW power. Thus, the exact value
of the quantized current $I_{0}=P_{l}n_{0}ef$ does not necessarily coincides
with the transfer of precisely integer number of electrons per period $n_{0}$, 
and the plateaus can move in parameter space.

In conclusion, we have shown that at low temperature, long-range Coulomb
interactions may have dramatic effect on the accuracy of the quantization of
the single-electron transport driven by surface acoustic waves through a
narrow constriction, formed in two-dimensional electron gas. The effect of
screening on the
SAW-induced potential near the edge of 2DEG can be described by a single
parameter - the time $\tau $ of the switching-off of the tunneling coupling
between 2DEG and the moving quantum dot. As a result, both the slope of the
plateaus and the width of the transition regions between the plateaus
saturate at low temperature to the values, determined by the characteristic
energy scale for nonadiabatic corrections $\hbar /\tau$.

We benefited from discussions with Henrik
Bruus, Yuri Galperin, Antti-Pekka Jauho, Anders Kristensen, and Julian
Shilton. This work was supported by EC under the SETamp project through the
contract SMT4-CT96-2049 (KF), through the contract SMT4-CT98-9030 (MP), by
the NSF under the Grants No. PHY94-07194 and DMR 9705406 (QN) and by Welch
Foundation (QN). Two of us (KF and QN) acknowledge the hospitality of ITP at
UC Santa Barbara, where part of this work was performed.

\end{multicols}


\begin{references}
\bibitem{experiments}  J. M. Shilton {\it et al.}, J. Phys.:Condens. Matter 
{\bf 8}, L531 (1996); V. I. Talyanskii {\it et al.}, Phys. Rev. B {\bf 56},
15180 (1997); J. Cunningham {\it et al.}, Phys. Rev. B {\bf 60}, 4850 (1999).

\bibitem{review}  See K. Flensberg {\it et al.}, 
Int. J. Mod. Phys. B {\bf 13}, 2651 (99) for a recent review of single electron metrology.

\bibitem{openchannel}  J. M. Shilton {\it et al.}, J. Phys.: Condens. Matter 
{\bf 8}, L337 (1996); H. Totland and Yu. Galperin, Phys. Rev. B {\bf 54},
8814 (1996).

\bibitem{thouless}  D. J. Thouless, Phys. Rev. B {\bf 27}, 6083 (1983); Q.
Niu, Phys. Rev. Lett. {\bf 64}, 1812 (1990); for a recent discussion, see B.
L. Altshuler and L. I. Glazman, Science {\bf 283}, 1864 (1999).
 
\bibitem{conform}  Due to the absence of translational
invariance in $y$-direction (see Fig.1), the problem is untractable 
analytically even with additional simplifying assumptions, such as 
made in 
L. I. Glazman and I. A. Larkin, Semicond. Sci. Techn. {\bf 6}, 32 (1991).

\bibitem{nakamura}  For comparison, in the recent state-of-the-art
experiments, fall/rise times of sharp voltage pulses used to
modulate the gate voltage are limited to $50\,{\rm ps}$: 
Y. Nakamura, Y. Pashkin, and J. Tsai, Nature {\bf 398}, 786 (1999).

\bibitem{AS}  A. Dorsey et al, Phys. Rev. B {\bf 40}, 3417 (1989); H. Shao,
P. Nordlander, and D. Langreth, Phys. Rev. Lett. {\bf 77}, 948 (1996).

\bibitem{pump}  C. Liu and Q. Niu, Phys. Rev. B {\bf 47}, 13031 (1993).

\bibitem{JWM}  A.-P. Jauho, N. Wingreen, and Y. Meir, Phys. Rev. B {\bf 50},
5528 (1994).

\bibitem{universality} To study the universality of the choice (\ref{V}), we use slightly different
expression, $V\left( t\right) =V_{0}\left( e^{t/2\tau }+1\right) ^{-1/2}$,
which coincides with (\ref{V}) for $t\gg \tau $. With this form of $V\left(
t\right) $, the $V_{0}$-independent result (\ref{finiteT}) is recovered, if $%
\Gamma _{0}=2\pi \nu V_{0}^{2}\gg 1/\tau $. For a special choice of
parameters, $\Gamma _{0}=1/\tau $, the result takes a form of (\ref
{finiteT}), but with $\cosh ^{2}$ instead of $\cosh $. This differs only
little from (\ref{finiteT}).


\bibitem{slope}  For instance, in the high temperature limit $T\gg 1/\tau $,
the results for the two-state and for the three-state approximations differ
only by factor of $2$: $S_{3}/S_{2}=2$.
\end{references}
\end{document}